\documentstyle[proceedings,psfig]{crckapb}
\def\spose#1{\hbox to 0pt{#1\hss}}

\def\kms{\ifmmode {\rm\,km\,s^{-1}}\else
    ${\rm\,km\,s^{-1}}$\fi}
\def\kmsMpc{\ifmmode {\rm\,km\,s^{-1}\,Mpc^{-1}}\else
    ${\rm\,km\,s^{-1}\,Mpc^{-1}}$\fi}

\def\msun{\ifmmode {\rm\,M_\odot}\else ${\rm\,M_\odot}$\fi}
\def\Msun{\ifmmode {\rm\,M_\odot}\else ${\rm\,M_\odot}$\fi}
\def\lsun{\ifmmode {\rm\,L_\odot}\else ${\rm\,L_\odot}$\fi}
\def\Lsun{\ifmmode {\rm\,L_\odot}\else ${\rm\,L_\odot}$\fi}
\def\rsun{\ifmmode {\rm\,R_\odot}\else ${\rm\,R_\odot}$\fi}
\def\Rsun{\ifmmode {\rm\,R_\odot}\else ${\rm\,R_\odot}$\fi}

\def\cm{{\rm\,cm}}
\def\cm3{\ifmmode {\rm\,cm^{-3}}\else ${\rm\,cm^{-3}}$\fi}

\def\ergps{\ifmmode {\rm\,erg\,s^{-1}}\else ${\rm\,erg\,s^{-1}}$\fi}
\def\ergpscm2{\ifmmode {\rm\,erg\,s^{-1}\,cm^{-2}}\else
    ${\rm\,erg\,s^{-1}\,cm^{-2}}$\fi}
\def\deg{\ifmmode {^{\circ}}\else {$^\circ$}\fi}
\def\degr{\ifmmode {^{\circ}}\else {$^\circ$}\fi}
\def\degs{\ifmmode {^{\circ}}\else {$^\circ$}\fi}

\def\etal{{\it et al.~}}

\def\h3Mpc{h^{-3}{\rm Mpc}^3}
\def\Ho{\ifmmode {\rm\,H_\circ}\else ${\rm\,H_\circ}$\fi}
\def\hnot{\ifmmode {\rm\,H_\circ}\else ${\rm\,H_\circ}$\fi}
\def\h0{\ifmmode {\rm\,H_\circ}\else ${\rm\,H_\circ}$\fi}
\def\hnotunit{\ifmmode {\rm\,km\,s^{-1}\,Mpc^{-1}}\else
    ${\rm\,km\,s^{-1}\,Mpc^{-1}}$\fi}

\def\qnot{\ifmmode {\rm\,q_\circ}\else ${\rm q_\circ}$\fi}
\def\q0{\ifmmode {\rm\,q_\circ}\else ${\rm q_\circ}$\fi}

\def\arcsec{\ifmmode {^{\prime\prime}~}\else $^{\prime\prime}~$\fi}
\def\asec{\ifmmode {^{\prime\prime}}\else $^{\prime\prime}$\fi}
\def\arcmin{\ifmmode {^{\prime}}\else $^{\prime}$\fi}
\def\amin{\ifmmode {^{\prime}}\else $^{\prime}$\fi}

\def\secper{\ifmmode \rlap.{^{s}}\else $\rlap{.}{^{s}} $\fi}
\def\minper{\ifmmode \rlap.{^{m}}\else $\rlap{.}{^m} $\fi}
\def\magper{\ifmmode \rlap.{^{m}}\else $\rlap{.}{^m} $\fi}
\def\arcsper{\ifmmode \rlap.{^{\prime\prime}}\else
    $\rlap.{^{\prime\prime}}$\fi}
\def\arcmper{\ifmmode \rlap.{^{\prime}}\else
    $\rlap.{^{\prime}}$\fi}
\def\spose#1{\hbox to 0pt{#1\hss}}
\def\simlt{\mathrel{\spose{\lower 3pt\hbox{$\mathchar"218$}}
     \raise 2.0pt\hbox{$\mathchar"13C$}}}
\def\simgt{\mathrel{\spose{\lower 3pt\hbox{$\mathchar"218$}}
     \raise 2.0pt\hbox{$\mathchar"13E$}}}

\begin{opening}
\title{MONSTERS AND BABIES FROM THE FIRST / IRAS SURVEY}
\author{Wil van Breugel}
\institute{IGPP/LLNL\\
P.O. Box 808, L-413, Livermore, CA 94550, USA}
\end{opening}
\runningtitle{The FIRST / IRAS Survey}
\begin{document}

\section{Abstract}

Radio continuum emission at cm wavelengths is relatively little affected
by extinction. When combined with far-infrared (FIR) surveys this
provides for a convenient and unbiased method to select (radio-loud)
AGN and starbursts deeply embedded in gas and dust--rich galaxies. Such
radio-selected FIR samples are useful for detailed investigations of
the complex relationships between (radio) galaxy and starburst activity, 
and to determine whether ULIRGs are powered by hidden quasars (monsters) 
or young stars (babies).

We present the results of a large program to obtain identifications and
spectra of radio-selected, optically faint IRAS/FSC objects using the
FIRST/VLA 20 cm survey (Becker, White and Helfand 1995). These 
objects are all radio-{\it `quiet'} in the sense that their radio
power / FIR luminosities follow the well-known radio/FIR relationship
for star forming galaxies.

We compare these results to a previous study by our group of a sample
of radio-{\it `loud'} IRAS/FSC ULIRGs selected from the Texas 365 MHz
survey (Douglas \etal 1996). Many of these objects also
show evidence for dominant, A-type stellar populations, as well as high
ionization lines usually associated with AGN. These radio-loud ULIRGs
have properties intermediate between those of starbursts and quasars,
suggesting a possible evolutionary connection.

Deep Keck spectroscopic observations of three ULIRGs from these samples
are presented, including high signal-to-noise spectropolarimetry.
The polarimetry observations failed to show evidence of a hidden quasar in
polarized (scattered) light in the two systems in which the stellar light
was dominated by A-type stars.  Although observations of a larger sample
would be needed to allow a general conclusion, our current data suggest
that a large fraction of ULIRGs may be powered by luminous starbursts,
not by hidden, luminous AGN (quasars).

While we used radio-selected FIR sources to search for evidence of a
causal AGN/starburst connection, we conclude our presentation with a
dramatic example of an AGN/starburst object from an entirely unrelated
quasar survey selected at the opposite, blue end of the spectrum.

\section{The FIRST/IRAS Survey: radio--quiet selected ULIRGs}

In order to investigate the AGN--starburst connection, particularly
at higher redshifts, we have constructed a new sample of potential
ultraluminous infrared galaxies (ULIRG) by making use of FIRST (``Faint
Image of the Radio Sky at Twenty cm''; Becker, White, and Helfand 1995)
survey which is currently underway at the VLA. This 1.4 GHz survey,
with unprecedented sensitivity ($\sim$1 mJy; 5$\sigma$) and resolution
(4$\arcsec$), will eventually cover $\pi$ steradians centered at the North
Galactic Cap.

We position matched the FIRST catalog with the $IRAS$ FSC catalog, and
selected the sources which follow the well known 
radio--FIR flux correlation for starburst galaxies and
which showed faint or no optical counterparts on the POSS.  We then used
the Lick 3m telescope to obtain optical identifications and spectroscopy
of the unknown sources. This resulted in a very high rate of finding
ULIRG at moderate redshifts. All except two 
of the approximately 70 sources observed
so far showed optical objects in our CCD imaging at the radio position.
Spectra with the KAST spectrograph yielded redshifts in the range $0.1
\le z \le 0.9$ and most of these are ULIRG at $z > 0.3$.

The new ULIRG in our FIRST/FSC (FF) sample are shown in Figure 1,
along with other representative classes of IR--luminous objects.  Their
radio--FIR properties place our FF sample mostly in the same region as the
original ULIRG sample of Sanders \etal (1988). The highest redshift
FF objects ($z = 0.710, 0.727, 0.904$) fall near the area in the
radio--FIR luminosity plane as other well--known ULIRGs. The FF optical
spectra contain typical starburst emission lines, including [O II],
[O III], and H$\alpha$ in the lower-$z$ objects, as well as absorption
lines characteristic of young stars. An example is plotted in Figure
2$a$ where a Keck spectrum of FF~J1614+3234 $z = 0.710$ shows [O II], the
4000 \AA ~break, Ca H+K, and several of the Balmer lines in absorption. 

\begin{figure}
\vspace{0cm}
\centerline{\psfig{file=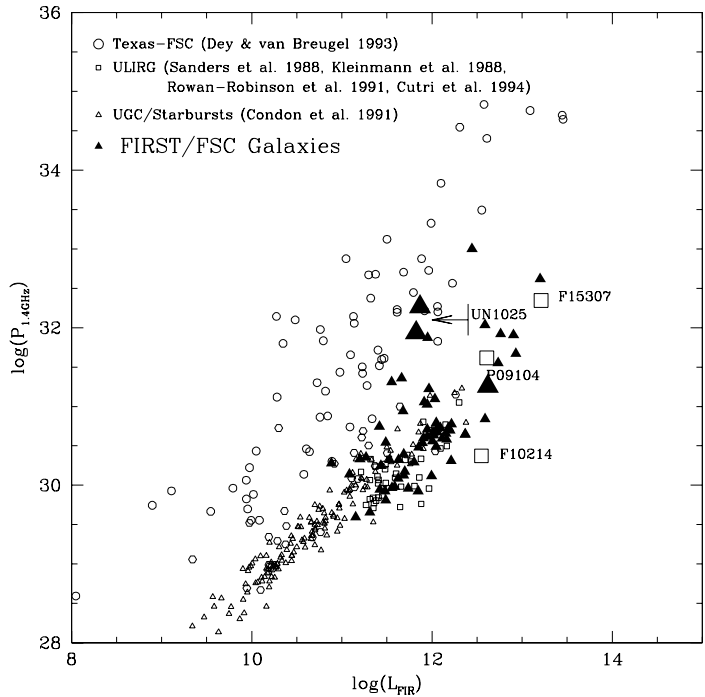,width=11cm}}
\caption{Radio vs. FIR luminosity plot for various radio selected IRAS/FSC
samples. The radio--quiet `starburst' track forms the bottom envelope,
while radio--loud galaxies and quasars form the upper envelope. The
region in between consists of intermediate--type objects, many of which
have stellar populations dominated by A-type stars. 
The large triangles 
refer to the three ULIRGs observed at Keck, as described in the text.}
\end{figure}

\section{The Texas/IRAS Survey: radio--loud selected ULIRGs}

A preliminary report about the Texas/IRAS results was given by Dey and
van Breugel (1993). The original sample was constructed by correlating
the Texas catalog with an early, pre-release version of the IRAS
catalog which included the Faint Source Catalog, as well as possible
spurious sources. One of the results from this original sample was the
possible identification of faint 60$\mu$ sources (3$\sigma$ - 4$\sigma$) 
with a number of high redshift far-infrared quasars.  However, the
far-infrared nature of these objects could not be confirmed by deep
mm-continuum or CO molecular line observations at the JCMT and IRAM 30m.
The radio/FIR/optical identification in these objects therefore remains
doubtful, or the faint IRAS detections are spurious. A sanitized version
of this sample, which excludes these objects, and includes only 
Texas/FSC (TF) sources, is shown in Figure 1. This sample remains of much
interest as it it shows the existence of a significant class of objects
which are intermediate between starburst galaxies and quasars. 

A large fraction of these intermediate luminosity systems show
deep Balmer absorption lines associated with relatively young, A-type
stellar populations similar to FF~J1614+3234 (Fig 2$a$), as well as high
ionization emission lines suggestive of the presence of AGN (but not
necessarily of quasar luminosities; see for example FF J1020+6436 in
Tran \etal 1999). They may form an important evolutionary link between
starburst and quasar activity.

\section{Keck spectropolarimetry of radio--selected ULIRGs}

FF~J1614+3234, with $L_{FIR} \sim 10^{12.6} \Lsun$, is one of the most
luminous ULIRG in our sample (we use the definition for L$_{FIR}$ 
as given by Sanders and Mirabel 1996). To determine the power source in this
and other ULIRGs we began a program to obtain high signal-to-noise
spectropolarimetry data of a number of galaxies with the Keck telescope.
The presence of a monster, possibly hidden in a dusty lair, might then
expected to be visible via indirect, reflected and hence polarized light.

We observed three ULIRGs in detail (Tran \etal 1999). Two ULIRGs with
dominant {\it young} ($A$-star type) stellar populations and weak high
ionization lines failed to show evidence for hidden quasars in polarized
(scattered) light. On the other hand, similar observations of a ULIRG with
only a modest young stellar population but with strong high ionization
lines did show polarized broad lines.  Other well--known examples in this
high ionization class which show evidence for polarized broad lines and
hidden quasars are the `hyper' luminous FIR galaxies P09104+4109 ($z$ = 0.44),
F15307+3252 ($z$ = 0.926), and F10214+4724 ($z$ = 2.286).

The detection of hidden quasars, using spectropolarimetry,
in this high ionization group but not in the low-ionization,
starburst-dominated ULIRGs (classified as LINERs or H II galaxies)
may indicate an evolutionary connection, with the latter being found in
younger systems. Since approximately 75\% of the FF objects in our sample
do not show any signs of high excitation emission lines the majority of
the ULIRGs may not contain monsters and even some of the most energetic
ULIRGs may be powered by massive starbursts (monstrous baby nurseries).

\begin{figure}
\vspace{0cm}
\centerline{\psfig{file=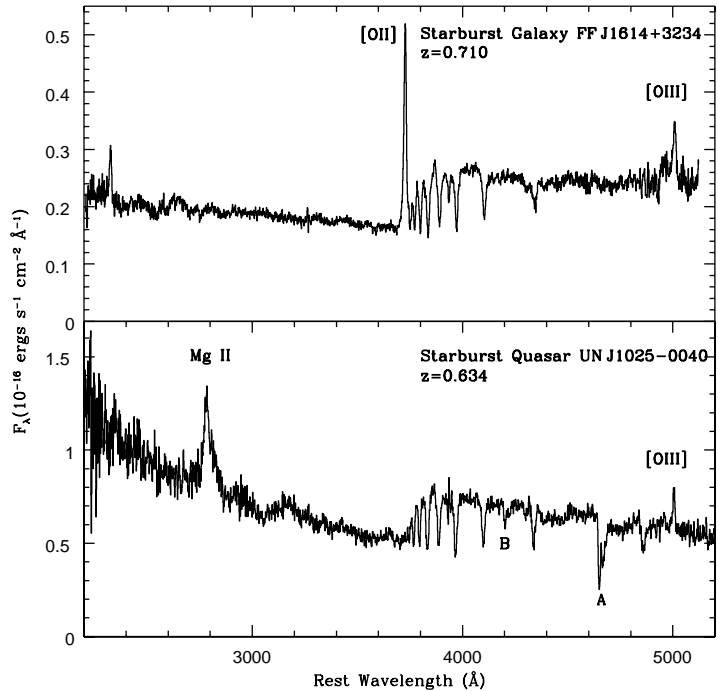,width=10cm}}
\caption{Keck spectra of the $L_{FIR} \sim 10^{12.6} \Lsun$ starburst
ULIRG FF J1614+3234 (top), and the
UV-excess starburst quasar UN J1025$-$0040.}
\end{figure}

\section{A Spectacular, UV-selected Starburst-Quasar} 

While our radio/far-infrared ULIRG project is specifically aimed to
search for a possible causal relationship between starburst and quasar
activity, our most spectacular `proto-type' object may have been found
accidentially during an entirely unrelated program aimed to study
radio-loud UV-excess quasars.

Using the NVSS survey (Condon \etal 1998) we selected radio loud quasars
from the 2dF quasar survey (Smith \etal 1998). We then observed a
subset of these using Keck `snapshot' spectroscopic observations and
discovered a spectacular `post-starburst quasar', UN~J1025$-$00400
($B=19$ and $z=0.634$; Brotherton \etal 1999).  The optical spectrum is
extraordinary, dominated by a quasar in the blue, and by a young, A-type
stellar population with a large Balmer jump and deep Balmer absorption
lines in the red (Fig 2$b$). There is no [OII] seen in emission, and
only a hint of broad H$\beta$.  Deep Keck spectropolarimetry showed weak
polarized continuum, but no strong polarized emission lines.

A Keck K-band image (0.5$\arcsec$ FWHM) fails to resolve the quasar
from the starburst, but does reveal surrounding asymmetric fuzz and a
nearby companion, suggestive of a galactic interaction.  Stellar synthesis
population models can reproduce the starlight component with a 400-Myr-old
instantaneous burst of 2$\times$ 10$^{10}$ \msun.  While starbursts and
interactions have been previously associated with quasars, no quasar
ever before has been seen with such a luminous young stellar population.

We searched the IRAS data base using ADDSCAN to determine whether
UN~J1025$-$00400 is also a far-infrared source. No emission was found at
60$\mu$ at a level of 0.15 Jy (3$\sigma$). The location of the source
in the radio/far-infrared luminosity diagram is indicated in Fig 1 and
does not rule out that UN~J1025$-$00400 may be a ULIRG and a member of
the intermediate class of objects between starbursts and quasars.

So, although we know a quasar is present, as well as a moderately aged
starburst, the low percentage polarization (1\% - 1.6\%) 
suggests that their geometry might be such that light from the quasar
nucleus and BLR does not intercept a large number of suitably placed
`reflectors'. This might arise, for example, if the starburst is located
in a plane (ring ?) orthogonal to the spin axis of the putative black
hole powering the quasar (and we are looking down this direction).
Alternatively the system may have little dust all together, and may not
be a ULIRG for that matter. Perhaps the starburst--quasar is a more
evolved system compared to the real intermediate class objects shown
in Fig 1. It's true location in this diagram might fall near the upper
envelope of the radio/FIR luminosity diagram.  Further far--infrared
observations of this system would be needed to investigate this.

% \section{Conclusions} : given in abstract

\noindent {\bf Acknowledgments} 

I thank the organizers for a most stimulating meeting, Alpen hike,
and Bier Fest fun, and my collaborators for allowing me to quote their
results in advance of publication.  Papers describing the observations in
greater detail are in preparation by S.A. Stanford, D. Stern, W. van
Breugel, C. De Breuck and A. Dey 1999 (on radio selected ULIRGs);
H. Tran, M.S. Brotherton, S.A. Stanford, W. van Breugel, A. Dey,
D. Stern and R. Antonucci 1999 (on Keck spectropolarimetry of ULIRGs);
and M.S. Brotherton, W. van Breugel, S.A. Stanford, R.J. Smith,
B.J. Boyle, A.V. Filippenko, L. Miller, T. Shanks, and S.M. Croom
1999 (on the UV-excess starburst quasar). The research at IGPP/LLNL
is performed under the auspices of the US Department of Energy under
contract W--7405--ENG--48.

\end{document}